\documentclass{iaus}
\usepackage{graphicx}
\def\aap{{\it A\&A}}
\def\apj{{\it ApJ}}
\def\apjs{{\it ApJS}}
\def\mnras{{\it MNRAS}}
\def\araa{{\it ARA\&A}}
\def\icarus{{\it Icarus}}

\title[Overview] 
{Organic Matter in Space - an Overview}

\author[van Dishoeck]   
{Ewine F. van Dishoeck$^{1,2}$
}

\affiliation{$^1$Leiden Observatory, Leiden University\\
P.O. Box 9513, NL--2300 RA Leiden, the Netherlands \\ email: {\tt ewine@strw.leidenuniv.nl}\\
$^2$ Max-Planck Institute f\"ur Extraterrestrische Physik, Garching, Germany
}

\pubyear{2008}
\volume{251}  
\pagerange{1--12}
\jname{Organic Matter in Space}
\editors{A.C. Editor, B.D. Editor \& C.E. Editor, eds.}
\begin{document}

\maketitle

\begin{abstract}
Organic compounds are ubiquitous in space: they are found in diffuse
clouds, in the envelopes of evolved stars, in dense
star-forming regions, in protoplanetary disks, in comets, on the
surfaces of minor planets, and in meteorites and interplanetary dust
particles.  This brief overview summarizes the observational evidence
for the types of organics found in these regions, with emphasis on
recent developments. The {\it Stardust} sample-return mission provides
the first opportunity to study primitive cometary material with
sophisticated equipment on Earth. Similarities and differences between
the types of compounds in different regions are discussed in the
context of the processes that can modify them.  The importance of
laboratory astrophysics is emphasized.
\end{abstract}

\firstsection 
\section{Introduction}

Organic matter is defined as chemical compounds containing
carbon-hydrogen bonds of covalent character, i.e., with the carbon and
hydrogen forming a true chemical bond. Observations over the last
century have established that these molecules are ubiquitous
throughout the universe, not only in our Galaxy (\cite{kwok07b}) but
even out to high redshifts (\cite{yan05}). With the detection of more
than 200 exoplanetary systems, a major question is whether these
organic compounds can be delivered in tact to new planetary systems
where they could form the basis for the origin of life. The answer
requires a good understanding of the entire lifecycle of organic
molecules from their formation in the outflows of evolved stars to the
diffuse interstellar medium (ISM) and subsequently through the
star-forming clouds to protoplanetary disks.  Ehrenfreund \& Charnley
(2000) and Ehrenfreund et al.\ (2002) summarized our understanding of
this cycle several years ago, but since then new observations (e.g.,
with the {\it Spitzer Space Telescope}), laboratory experiments, and
in-situ space missions (in particular the {\it Stardust} mission)
have occurred. Thus, a symposium such as this reviewing these recent
developments is timely.

In this opening paper, a broad overview will be given where organic
matter is found in space and which species have been identified.
Subsequently, observational evidence of how and where these molecules
are modified will be summarized. Finally, a number of questions to
address at the symposium and in the future are raised.  The importance
of laboratory astrophysics in providing the basic data to interpret
astronomical and solar-system observations and analyse meteoretic and
cometary samples is emphasized.

\section{The need for laboratory astrophysics}

Several decades ago, Henk van de Hulst accused cosmologists of
`playing tennis without a net', when they were putting forward many
models that could not be tested by any observations.  Similarly, much
of astrochemistry (and, in fact, much of astronomy as a whole) would
be `playing tennis without a net' if there were no laboratory data
available to analyse and interpret the observational data of
astronomical sources. The list of required data for organic compunds
is extensive, and getting such information for even a single molecule
often involves the building up of sophisticated laboratory equipment
followed by years of painstaking data taking.

The most basic required information is spectroscopy of organic
molecules from UV to millimeter wavelengths to identify the sharp
lines and broad bands observed toward astronomical sources. One recent
development in this area is the use of cavity ringdown spectroscopy to
increase the sensitivity compared with classic absorption spectroscopy
by orders of magnitude, which has allowed measurements of rare species
that can be produced only in small amounts. Examples include gaseous
polycyclic aromatic hydrocarbons (PAH) (e.g.,
\cite{salama05,huisken07}) or carbon chains (e.g.,
\cite{dzhonson07,linnartz00}), in addition to matrix-isolation studies
of large samples of PAHs (e.g., \cite{hudgins99}). Spectroscopy data
bases of solids, including silicates (e.g., Jaeger et al.\ 1998,
2003), carbonates (e.g., \cite{posch07}), ices (e.g.,
\cite{bisschop07c,bernstein05}) and carbonaceous material (e.g.,
\cite{mennella97,jaeger06,munoz06}) continue to grow.

The next step in understanding organics is to obtain rates for the
various reactions that are expected to form and destroy organics under
space conditions. Here recent developments include measurements and
theory of gaseous neutral-neutral rate coefficients at low
temperatures (e.g., \cite{chastaing01,smith06}), branching ratios for
dissociative recombination (e.g., \cite{geppert04}), and rates for
photodissociation of molecules exposed to different radiation fields
(\cite{vanhemert08}). Surface science techniques at ultra high-vacuum
conditions are now being applied to study thermal- and
photo-desorption (e.g., \cite{collings04,oberg07}) and formation of
simple organic ices at temperatures down to 10~K (e.g.,
\cite{watanabe04,bisschop07a}), while more traditional set-ups
continue to provide useful information on the formation of complex
organics in ices exposed to UV (e.g., \cite{elsila07,munoz03}) and to
higher energy particle bombardment (e.g., \cite{moore00}). There is
also a wealth of new literature on the formation of carbonaceous
material in discharges (e.g., \cite{imanaka04}) and its processing at
higher temperatures and when exposed to UV (e.g., \cite{dartois07}).

Finally, the techniques to analyze meteoritic and cometary material in
the laboratory have improved enormously in the last decade, and now
allow studies of samples on submicrometer scale. Examples include
ultra-L$^2$MS and nano-SIMS (e.g., \cite{messenger07}), XANES (e.g.,
\cite{flynn06}) and NMR (e.g., \cite{cody05}). Their development was
essential for analysis of the samples returned by {\it Stardust}.

\begin{figure}{t}
\begin{center}
\includegraphics[scale=.35,angle=-90]{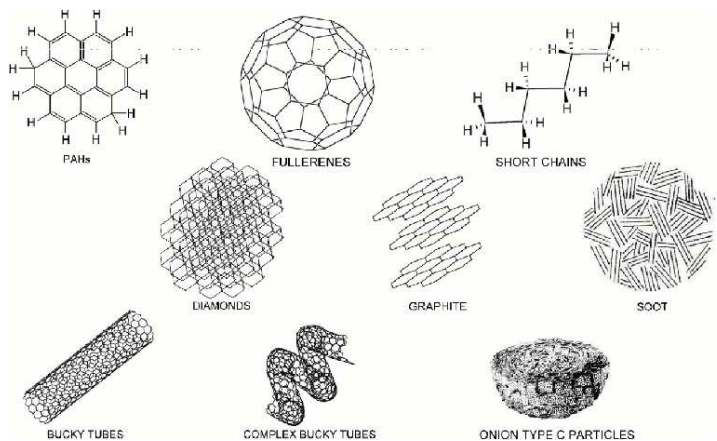}
\end{center}
\caption{Examples of different types of carbonaceous material which are
likely present in the ISM and solar system
(from: \cite{ehren00})}.
\end{figure}

\section{Which organic compounds are found where?}

In the following sections, the observational evidence for organic
material is summarized, together with the identification of the type
of material, where possible (Fig.\ 1). For small gas-phase molecules,
the identification is unambiguous, but for larger compounds often only
the types of carbon bonds making up the material can be specified.
Carbon can be bonded in several ways: a triple CC bond with single H
on the side (e.g., HC$\equiv$CH, denoted as $sp$ hybridization); a
double CC bond with two H's on each side (e.g, H$_2$C=CH$_2$, denoted
as $sp^2$ hybridization) and a single CC bond with three H's on each
side (e.g., H$_3$C-CH$_3$, denoted as $sp^3$ hybridization). In
aromatic material, the electrons are delocalized over the entire
molecule such as in a benzene ring or a polyethylene chain with
alternating double and single CC bonds.  The molecule thus contains
$sp^2$ bonds.  In aliphatic material, no double bonds occur, and only
$sp^3$ hybridization is found. Spectral signatures of organic
compounds include strong electronic transitions at optical and UV
wavelengths, vibrational transitions at infrared wavelengths, and pure
rotational transitions at millimeter wavelengths.  In the latter case,
the molecule needs to have a permanent dipole moment to be detected.

The majority of carbon in interstellar clouds (at least 50\%) is in
some form of carbonaceous solids with grain sizes large enough
($\sim$0.1 $\mu$m) not to have any clear spectroscopic signature,
other than continuous opacity. Another fraction of the carbon (up to
30\%) can be in gaseous C, C$^+$ and/or CO, or in CO and CO$_2$
ices. Most of the discussion in this paper concerns the remaining
$\sim$20\% of carbon present in carbonaceous molecules, ices and small
grains.

\subsection{Diffuse and translucent clouds}

Diffuse and translucent molecular clouds are concentrations of the
interstellar gas with extinctions up to a few mag (for review, see
\cite{snow06}). Typical temperatures range from 15--80 K and densities
from a few 100--1000 cm$^{-3}$. These are the only types of clouds for
which high quality optical and UV spectra can be obtained by measuring
the electronic transitions in absorption against bright background
stars. In addition to the simplest organics CH, CH$^+$ and CN detected
in 1937--1941, a series of more diffuse features called the 'diffuse
interstellar bands' (DIBs) has been known since 1922. Nearly 300 DIBs
are now known in the 4000--10000 \AA \ range (e.g., \cite{hobbs08}),
but not a single one has yet been firmly identified despite numerous
suggestions by the world's leading spectroscopists.  Two bands at 9577
and 9632 \AA \ are consistent with features of C$_{60}^+$, but
laboratory spectroscopy of gaseous C$_{60}^+$ is needed for firm
identifcation of the first fullerene in interstellar space
(\cite{foing97}). Long carbon chains with $n<10$ and small PAHs have
been excluded (\cite{maier04,ruiterkamp05}), but larger versions are
possible.  The amount of carbon locked up in the DIB carriers is
likely small, $<1\%$, assuming typical oscillator strengths for large
molecules.

Measurements of the UV extinction curve show a bump at 2175 \AA,
characteristic of the $\pi \to \pi^*$ transition in carbonaceous
material. The precise identification is still uncertain, with both
graphitic and hydrogenated amorphous carbon (HAC, a material
consisting of islands of aromatic C joined by a variety of peripheral
$sp^2$- and $sp^3$-bonded hydrocarbons) leading candidates (e.g.,
\cite{draine03}). A major puzzle is why no DIB features have yet been
seen at UV wavelengths, since small carbon-bearing molecules should
have strong electronic transitions in this region.

\begin{figure}
\begin{center}
\includegraphics[scale=.35,angle=-90]{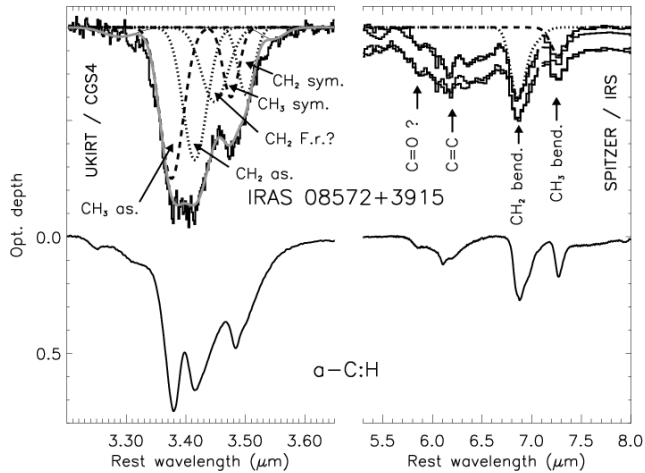}
\end{center}
\caption{$Spitzer$ and ground-based IR spectrum toward the galaxy IRAS
  08572+3915 compared with a laboratory spectrum of amorphous carbon
  a-C:H. The spectrum in the 5--8 $\mu$m range excludes large amounts
  of O- and N- containing groups. It suggests a larger ratio of
  aliphatic over aromatic content than thought before
  (\cite{dartois07})}.
\end{figure}

At IR wavelengths, the characteristic `unidentified infrared bands'
(UIR) bands are seen in emission throughout the diffuse ISM, even in
clouds exposed to the normal interstellar radiation field (see review
by \cite{tielens08}). The commonly accepted identification is with
PAHs of sizes small enough ($N_C\approx$50) to be excited by UV and
emit in discrete bands. It is clear that the fraction of O and N in
PAHs is very low, but a small amount of N (few \% w.r.t.\ carbon) has
been proposed to explain the small shift of the interstellar 6.2
$\mu$m feature compared with laboratory data of pure PAHs
(\cite{hudgins05}). Alternative explanations include the presence of
small side chains (\cite{sloan05}). The fraction of carbon locked up
in small PAHs is estimated to be $\sim$4\%. Large PAHs or PAH clusters
may be responsible for the plateaus underlying the discrete PAH bands,
containing of order 2\% of the carbon, whereas the Very Small Grains
(VSGs) responsible for the 12 $\mu$m IRAS emission contain a similar
amount (\cite{tielens08}).

In lines of sight which sample large columns of diffuse gas (e.g.,
toward the Galactic center, external galaxies), a feature at 3.4
$\mu$m has been seen (see review by \cite{pendleton04}).  This
wavelength is characteristic of the stretching modes of -CH$_2$ and
-CH$_3$ groups in aliphatic material (\cite{duley83}). Many materials
have been proposed, ranging from HAC to Quenched Carbonaceous
Composites (QCC), Kerogen (a coal-like material consisting of arrays
of aromatic carbon sites, aliphatic chains and linear chains of
benzenic rings) and photoprocessed carbon-containing ices. New
constraints come from {\it Spitzer} observations of the 5--8 $\mu$m
region (\cite{dartois07}) (Fig.\ 2). The absence of strong CO and CN
bands points again to little N and O, but the -CH$_2$ and -CH$_3$
bending modes are clearly seen and can be well fitted by a HAC or
a-C:H (amorphous carbon) material, which typically contains about 15\%
of the available carbon. A possible representation of the carrier of
the 3.4 $\mu$m feature is presented in Fig.\ 3.

Finally, small carbonaceous molecules like C$_3$H$_2$ have been
detected in absorption at mm wavelengths toward distant
quasars whose line of sight passes through a galactic molecular cloud
(\cite{lucas00}). These molecules are also seen in emission in
Photon-Dominated Regions (PDRs) with abundances that are much higher
than expected from standard ion-molecule reactions. One possible
explanation is that they result from photodissociation of larger
carbonaceous molecules such as PAHs (\cite{pety05}).

\begin{figure}
\begin{center}
\includegraphics[scale=.35,angle=-90]{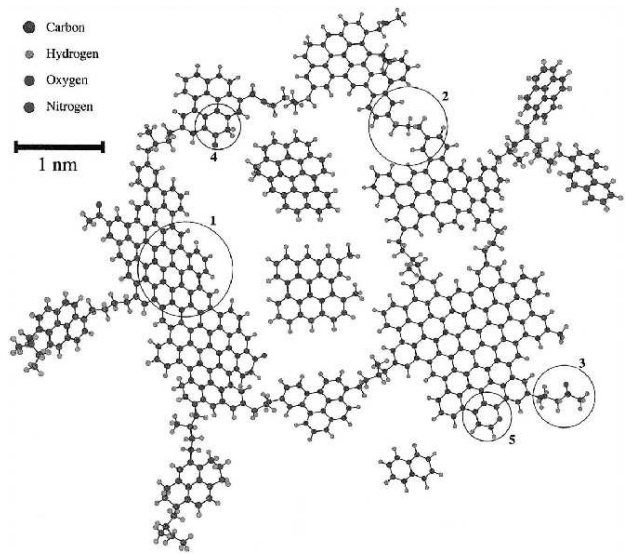}
\end{center}
\caption{Proposed structure of the carbonaceous interstellar dust in
  the diffuse ISM by Pendleton \& Allamandola (2002).  The structure
  is a kerogen-type aromatic network bridged by aliphatic chains,
  including side groups and hetero-atoms. A typical 0.1 $\mu$m
  carbonaceous grain would contain $\sim 10^4$ of these fragments. One
  of the challenges for this meeting is to determine whether this
  picture is still valid, or whether, for example, more aliphatic
  chains have to be added.}
\end{figure}

\subsection{Evolved stars}

The spectra of carbon-rich (proto-)planetary nebulae (PPN) are full of
IR emission features, with both the aromatic PAH and the aliphatic 3.4
$\mu$m bands commonly observed in emission (e.g.,
\cite{goto07}). Interestingly, the aromatics are stronger and
aliphatics weaker in planetary nebulae (PN) compared with earlier
evolutionary stages. Also, they appear to be larger than in the
diffuse ISM, with $N_C\approx 100-200$. In the AGB phase leading up to
the extreme carbon stars, C$_2$H$_2$ absorption at 13.7 $\mu$m (the
building block to make benzene) and emission features at 21 and 30
$\mu$m are seen, with the latter two still unidentified (see overview
by \cite{kwok07a}).

A particularly interesting object is the Red Rectangle, a
protoplanetary nebula which shows prominent Extended Red Emission
(ERE), both as a continuum at 5000--7000 \AA\ and in discrete bands
(\cite{winckel02}). Some of the latter are close to the positions of
strong DIBs seen in diffuse clouds. The positions and shapes of both
the IR and optical bands may vary with distance from the star,
suggesting a change in the composition of the material with UV dose
(\cite{song03}).

\subsection{Dense star-forming regions}

Cold dense cores are the realm of the long unsaturated carbon chains
such as HC$_9$N, discovered in the 1970's. Recent developments include
the identification of negative ions such as C$_6$H$^-$ and C$_8$H$^-$
(\cite{mccarthy06}) as well as more saturated chains such as
CH$_2$CHCH$_3$ (\cite{marcelino07}). Taken together, these chains make
up only a small fraction, $<$0.1\%, of the total carbon budget.

Saturated complex organic molecules such as CH$_3$OH, CH$_3$OCH$_3$
and C$_2$H$_5$CN are commonly seen in high abundances toward warm
star-forming regions such as Orion and SgrB2, which have been surveyed
at (sub)mm wavelengths for more than 30 years (e.g.,
\cite{schilke01}). Such `hot cores' have been detected around most
massive protostars and are now commonly used as a signpost of the
earliest stages of star formation. One recent development is that they
are also found around low-mass protostars, with IRAS 16293--2422 as
the prototypical example (\cite{cazaux03}).  Abundance ratios from
source to source are remarkably constant (e.g.,
\cite{bisschop07b,bottinelli07}) pointing to an origin in grain
surface chemistry, although some variations between low- and high-mass
sources are found. Also, a clear segregation of oxygen- and
nitrogen-rich organics is seen (e.g., \cite{wyrowski99}).  One of the
major questions is whether all observed complex organics are produced
in the ice or whether some of them are formed in the hot gas following
evaporation of ices (\cite{charnley92}). Each organic molecule has an
abundance of typically $10^{-9}-10^{-7}$ with respect to H$_2$, but
the total fraction of carbon locked up in these complex molecules can
amount to a few \%.

More complex organic molecules such as amino acids and bases, which
are relevant for pre-biotic material, have not yet firmly been
identified. Indeed, the spectra of hot cores are so crowded that line
confusion is a serious issue. Ethylene glycol, CH$_2$OHCH$_2$OH, a
complex organic found in comets, has been claimed in SgrB2
(\cite{hollis02}), but its detection is not yet fully secure.

The largest reservoir of volatile carbonaceous material is in the
ices, whose strong mid-infrared absorption bands are seen not only
toward most massive protostars (\cite{gibb04}) but also toward a wide
variety of low-mass YSOs (e.g., \cite{boogert08}). Besides H$_2$O ice,
CO, CO$_2$, OCN$^-$, CH$_4$, HCOOH, CH$_3$OH and NH$_3$ ice have been
identified. The recent surveys of low-mass YSOs show that some
molecules like CH$_4$ have relatively constant abundances of $\sim$5\%
with respect to H$_2$O ice (\cite{oberg08}), whereas those of CH$_3$OH
vary from $<1$ to 25\% (see Bottinelli et al., this
volume). Altogether, the known organic molecules (excluding CO and
CO$_2$) may lock up to 10\% of the available carbon.

Most significant is the absence of PAH and 3.4 $\mu$m emission or
absorption in the cold cores and deeply embedded stages of star
formation. Indeed, a recent {\it Spitzer} and VLT survey of low-mass
embedded YSOs shows no detections, indicating PAH abundances that are
at least a factor of 10 lower than in the diffuse gas, perhaps due to
freeze-out (\cite{geers08}).  No absorptions due to PAHs in ices have
yet been found, but the lack of basic spectroscopy prevents
quantitative limits.

\subsection{Protoplanetary disks}

Once the collapsing cloud has been dissipated, a young star emerges
which can be seen at visible wavelengths but is still surrounded by
a protoplanetary disk. PAH emission has been detected in roughly half of
the disks surrounding Herbig Ae stars, i.e., intermediate mass young
stars (\cite{acke04}). More recently, PAHs have also been seen in a
small fraction ($\sim 10\%$) of disks around solar mass T Tau stars
(\cite{geers06}). A quantitative analysis of the emission indicates
PAH abundances that are typically factors of 10--100 lower than in the
diffuse ISM, either due to freeze-out or caused by coagulation.  The
spatial extent of the PAH emission measured with adaptive optics on 8m
class telescopes is of order 100 AU, i.e., comparable with the size of
the disk, but varying with feature (\cite{habart04}).  Modeling of the
spatial extent as well as the destruction of PAHs by the intense UV or
X-ray emission from the star indicates that the PAHs must be large,
$N_C\approx 100$ (\cite{geers07a,visser07}).

Smaller organics are present in high abundances in the inner disk
($<10$ AU).  Indeed, hot (400-700 K) C$_2$H$_2$ and HCN have been
detected in absorption in edge-on disks with abundances factors of
1000 larger than in cold clouds (\cite{lahuis06}). Recently, they have
also been seen in emission (\cite{carr08}). The observed abundances
are consistent with models of hot dense gas close to LTE (e.g.,
\cite{markwick02}).

A particularly intriguing class of disks is formed by the so-called
transitional or `cold' disks with large inner holes. An example is Oph
IRS 48, in which a large (60 AU radius) hole is revealed in the large
grain 19 $\mu$m image. Interestingly, PAHs are present inside the
hole, indicating a clear separation of small and large grains in
planet-forming zones (\cite{geers07b}).  Another intriguing case is
formed by the more evolved disk around HR 4796A, which has likely lost
most of its gas and is on its way to the debris-disk stage. Recent HST
imaging shows colors with a steep red slope at 0.5--1.6 $\mu$m and
subsequent flattening off (\cite{debes08}) (Fig.\ 4). These colors are
reminiscent of those of minor planets in our solar system, such as the
Centaur Pholus, where the data are best fitted with tholins, i.e.,
complex organics produced in the laboratory in a CH$_4$/N$_2$
discharge with characteristics similar to Titan's haze (e.g.,
\cite{cruikshank05}).

\begin{figure}
\begin{center}
\includegraphics[scale=0.35,angle=0]{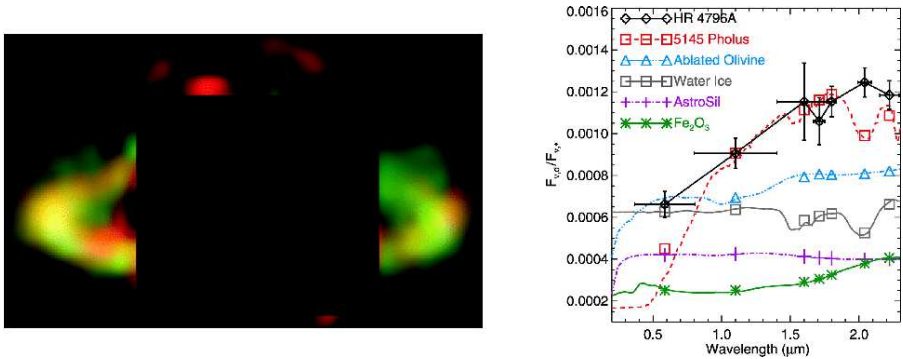}
\end{center}
\caption{Left: False color HST image ($\sim 4''\times 4''$) of the HR
  4796A disk. Blue corresponds to 0.58 $\mu$m, green to 1.10 $\mu$m
  and red to 1.71 $\mu$m.  Right: Disk / stellar flux ratio as
  function of wavelength. For comparison, grain models for candidate
  materials with a $n^{-3.5}$ size distribution with $a_{\rm min}$=3
  $\mu$m and $a_{\rm max}=$1000 $\mu$m are shown, normalized to the
  1.10 $\mu$m data for HR 4796A and offset for clarity (from:
  \cite{debes08}).}
\end{figure}

\subsection{Comets and minor planets}

Many volatile organics have been detected in bright comets like
Hale-Bopp thanks to improved sensitivity at IR and mm wavelengths (see
reviews by Bockel\'ee-Morvan et al.\ 2000, 2006). Most of them are
parent species evaporating directly from the ices. The list includes
HCN, C$_2$H$_2$, C$_2$H$_6$, CH$_3$OH, .... , all of which except
C$_2$H$_6$ have also been detected in star forming regions in the ice
or gas. Typical abundances are 0.1--few \% with respect to H$_2$O ice.
A larger variety of comets originating from both the Oort cloud and
Kuiper Belt have now been sampled, and variations in abundances
between comets are emerging, with organics like CH$_3$OH and
C$_2$H$_2$ depleted by a factor of 3 or more in some comets (e.g.,
\cite{kobayashi07}). PAHs have not yet been firmly identified by
ground-based 3.3 $\mu$m spectra.

Fly-bys through the comae of Comets Halley, Borelly and Wild-2 have
provided a much closer look at cometary material, including in-situ
mass spectrometry of the gases.  A major discovery of the {\it Giotto}
mission to Halley was the detection of the so-called CHON particles:
complex, mostly unsaturated, organics with only a small fraction of
O and N atoms (\cite{kissel87}).

A major question is whether the evaporating gases are representative
pristine material unchanged since the comets were formed more than 4
billion yr ago, or whether they have been changed by `weathering'
(e.g., high-energy particle impact) during their long stay in the
outer solar system.  The {\it Deep Impact} mission to Comet Tempel 1
was specifically designed to address this question, by liberating
pristine ices from deep inside the comet following impact
(\cite{ahearn05}). A sigificant increase in the IR emission around 3.5
$\mu$m, characteristic of CH$_3$CN and CH-X bands was seen immediate
after impact, but no strong PAH bands were evident.

\begin{figure}{t}
\begin{center}
\includegraphics[scale=.35,angle=180]{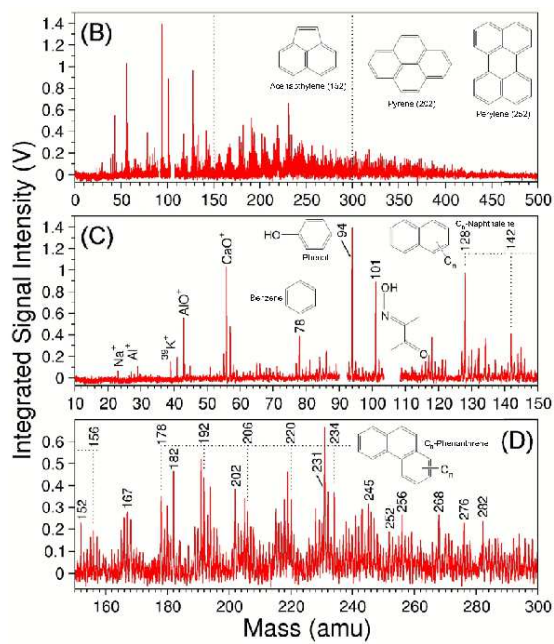}
\end{center}
\caption{$\mu ltra$-L$^2$MS analysis of one of the $Stardust$ samples,
  showing the presence of small PAHs (from: \cite{clemett07}).}
\end{figure}

The {\it Stardust} mission has taken a major step forward in the study
of primitive solar system material by returning samples from Comet
Wild-2 back to Earth, where they can be subjected to in-depth
laboratory analysis using the most sophisticated experimental
techniques (\cite{brownlee06}). Wild-2 is a less evolved comet than
others, having spent most of its lifetime in the Kuiper Belt and being
captured into its currrent orbit only 30 years ago. Thus, it should
not have suffered much thermal heating close to the Sun. Many complex
organic molecules are found in the analysis of the {\it Stardust}
particles to date, with a heterogeneous distribution in abundance and
composition between particles. Many of the organics are PAHs, with
typical sizes of just a few rings, i.e., generally smaller than the
PAH size inferred in the ISM (\cite{sandford06,clemett07})
(Fig.\ 5). Also, a new class of aromatic poor organic material is
found compared with those seen in IDPs and meteorites, perhaps related
to the fact that Wild-2 has had less thermal processing. The material
appears richer in O and N than meteoritic organics. A major challenge
for future studies will be to quantify the organics produced by the
particle impacts inside the aerogel and isolate those from true
cometary material.

Evidence for the presence of organics on other minor planets comes
from their red colors at optical and near-IR wavelengths as seen
in reflected sunlight. A particularly well studied case is the surface
of the Centaur object Pholus (\cite{cruikshank05}). Since discrete
spectral features are lacking, identification of the material is not
unique, but energy deposition in gas and ice mixtures containing
CH$_4$ and N$_2$ produces tholins with colors
similar to those observed (\cite{imanaka04}).

Titan is particularly interesting because its atmosphere is thought to
be similar to that of our (primitive) Earth, with the main difference
being that it consists mostly of N$_2$ and CH$_4$ rather than N$_2$
and CO$_2$. The {\it Cassini} mission has studied Titan's haze in
detail and the descent of the {\it Huygens} probe through the
atmosphere has indeed revealed many nitrogen-rich organics
(\cite{niemann05}). Methane in the atmosphere must be continuously
replenished by cryo-volcanism or other processes, since its lifetime
due to photochemistry is short.

Besides tholins, HCN polymers have also been speculated to be part of
the dark component present on outer solar system bodies, including
comets (\cite{matthews06}). It can also contribute to the orange haze
in the stratosphere of Titan. Overall, it is clear
that organics are a widespread component of solar system material.

\subsection{Meteorites and IDPs}

The most primitive and least processed meteorites ---the so-called
carbonaceous chondrites --- contain ample organic material. Well known
examples are the Murchison, Orgueil and Tagish Lake meteorites, which
contain up to 3\% by weight in carbon-rich material. Most of the
organics (60-80\%) are in an insoluable macromolecular form, often
described as `kerogen-like'. The remaining 20\% are in soluable form
and have been found to contain corboxylic acids, PAHs, fullerenes,
purines, amides, amides and other pre-biotic molecules (e.g.,
\cite{cronin93,botta02}). Amino acids --more than 80 different types--
have also been found, but are likely formed from reactions of liquid
water with HCN and H$_2$CO under the high pressure in the parent body
rather than being primitive solar system material.

Interplanetary dust particles (IDPs) have been collected through
stratospheric flights over the past decade and analyzed in detail in
the laboratory.  Organic carbon, including aliphatic hydrocarbons and
the carrier of the 2175 \AA\ feature, are common
(\cite{flynn00,bradley05}).

\section{Evolution of organic matter}

As organic material evolves from the evolved stars to the diffuse and
dense ISM, and subsequently from collapsing envelopes to disks, icy
solar sytem bodies and meteorites, many processes can affect their
composition and abundance (\cite{ehren06}). From the AGB and PPN phase
to the PN phase, UV processing changes aliphatics to aromatics
material (\cite{kwok07a}).  In the subsequent step, the organics can
be shattered by shocks as they enter the diffuse ISM and are exposed
to passing shocks from supernovae and winds
(\cite{jones96}). Destruction of graphite produces very small
carbonaceous grains, including presumably the smaller PAHs. When the
organics enter the dense cloud phase, freeze-out will affect all
organics, coagulation can occur, and the changing balance between UV
dissociation and re-hydrogenation can modify the aliphatic to aromatic
abundance ratio (e.g., \cite{munoz01}). Also, volatile complex
organics are formed on the grains as ices.

Once the molecules enter the inner part of the collapsing envelope,
ices will evaporate and some of the material is transported into
disks, either as ice or gas. UV radiation and heating can further
process the material before it becomes incorporated into cometary or
planetary material. Here weathering and processing over the last 4.5
billion yr can change the top layers of the parent body, whereas
aqueous alteration and thermal processing can further change the
composition of organics in meteorites. Given all this potential
processing, it is natural to expect the organics in the different
sources to vary substantially.

\subsection{Similarities}

One of the strongest pieces of evidence that some organic material may
remain unaltered through this entire cycle comes from the similarity
of the 3.4 $\mu$m feature in the diffuse ISM and in meteorites
(\cite{pendleton04}).  Also, the mere presence of PAHs in
(proto)planetary nebulae, the diffuse ISM, comets and meteorites
suggests that these molecules are not fully destroyed during the
lifecycle from evolved stars to solar systems.

Other intriguing piece of the puzzle is provided by the similarity in
the composition and abundances of ices in protostellar regions and
comets: is this just a coincidence or are the original ices preserved
as they enter the comet-forming regions? The answer likely depends on
the volatility of the species. Also, the similar red colors found in
at least one protoplanetary disk and those seen on minor planets hint
at preservation of the more refractory organics in the transport
through disks. Finally, the presence of the 2175 \AA\ bump in IDPs and
in the ISM suggests a common carbonaceous carrier.

\subsection{Differences}

There is also abundant evidence that not all organics seen in the
various sources are the same. The PAH and 3.4 $\mu$m aliphatic
features are not seen in dense clouds or protostellar sources, with
abundances of the carriers inferred to be lower by factors of 10--100,
most likely due to freeze-out. Also, the sizes of PAHs in meteorites
are smaller than those in the ISM and disks, which in turn are smaller
than those found in (proto-)planetary nebulae.  Aliphatic material is
transformed into aromatics with increasing UV dose in evolved stars.
Thus, most of the PAHs seen in protoplanetary disks are not the same
as those seen around evolved stars.

In terms of volatiles, some large organics are clearly much more
abundant in comets than in interstellar gases or ices, with ethylene
glycol a good example. Organics in comets are also different compared
with IDPs and meteorites, both in terms of PAHs and other
species. This suggests processing during planetary formation or during
the journey of the meteorites and IDPs to Earth.

\section{Some open questions}

The above summary raises many questions that shoud be addressed in
this meeting and future studies.  How is carbonaceous material formed
in the envelopes of evolved stars and how does the composition depend
on stellar parameters? What is the main form of solid carbon? How
important is UV processing in modifying organic matter, or even
producing it (e.g., from UV processing of water-poor ices)?  How
relevant are the volatile complex organics found in star-forming
regions to the origin of life? Do they survive the transport from
clouds to disks to planetary systems? What is the link between
interstellar and solar system refractory macromolecular carbon? Where
do the CHON particles found in comets fit in? How are the interstellar
PAHs modified in disks, and where is the `soot' line in relation to
the terrestrial planet-forming zones?

The future is bright thanks to upcoming new observational facilities
such as ALMA, further development of infrared interferometers,
HST-COS, JWST and ELTs. Complementary laboratory work will be even
more essential to make progress.

The author or grateful to S.\ Kwok for the invitation and hospitality.
This work is supported by a Spinoza grant from NWO.

\end{document}